\documentclass[12pt,preprint]{aastex}
\begin{document}

\title{IRAS 16293-2422: Evidence for Infall onto a Counter-Rotating Protostellar Accretion Disk}

\author{Anthony J. Remijan\altaffilmark{1,2} and J. M. Hollis\altaffilmark{1}}

\altaffiltext{1}{NASA Goddard Space Flight Center, Computational and Information Sciences and Technology Office, Code 606, Greenbelt, MD  20771}
\altaffiltext{2}{National Research Council Resident Research Associate}

\begin{abstract}

We report high spatial resolution VLA observations of the low-mass
star-forming region IRAS 16293-2422 using four molecular probes: 
ethyl cyanide (CH$_3$CH$_2$CN), methyl formate (CH$_3$OCHO), formic acid (HCOOH), 
and the ground vibrational state of silicon monoxide (SiO).  Ethyl cyanide emission has a
spatial scale of $\sim$20$''$ and encompasses binary cores A and B as
determined by continuum emission peaks.  Surrounded by formic acid
emission, methyl formate emission has a spatial scale of  $\sim$6$''$and is
confined to core B.  SiO emission shows two velocity
components with spatial scales less than  2$''$ that map $\sim$2$''$ northeast of
the A and B symmetry axis.  The redshifted SiO is $\sim$2$''$ northwest of
blueshifted SiO along a position angle of $\sim$135$^o$ which is
approximately parallel to the A and B symmetry axis.  We interpret the
spatial position offset in red and blueshifted SiO emission as due to
rotation of a protostellar accretion disk and we derive $\sim$1.4 M$_{\odot}$ 
interior to the SiO emission.  In the same vicinity, Mundy et
al.\ (1986) also concluded rotation of a nearly edge-on disk from OVRO
observations of much stronger and ubiquitous $^{13}$CO emission but the
direction of rotation is opposite to the SiO emission findings. 
Taken together, SiO and $^{13}$CO data suggest evidence for a
counter-rotating disk.  Moreover, archival BIMA array $^{12}$CO data show
an inverse P Cygni profile with the strongest absorption in close
proximity to the SiO emission, indicating unambiguous material infall
toward the counter-rotating protostellar disk at a new source location within
the IRAS 16293-2422 complex.  The details of these
observations and our interpretations are discussed.

\end{abstract}

\keywords{ISM: abundances - ISM: clouds - ISM: individual (IRAS 16293-2422) - ISM: molecules - radio lines:  ISM}

\section{INTRODUCTION}

IRAS 16293-2422 is a low-mass star forming region located in the $\rho$ Ophiuchus 
cloud complex at a heliocentric distance of 160 pc.  It contains an undetermined number of protostellar
objects, high velocity outflows in the E-W and NE-SW directions (Stark et al.\ 2004), and the region 
is dominated by two radio continuum peaks designated cores A and B. These two principal cores
are separated by $\sim$5$''$ and are often referred to as the ``binary'' system with
a mass of 0.49 M$_{\odot}$ for core A and 0.61 M$_{\odot}$ for core B (Looney, Mundy \& Welch 2000).  
Moreover, the gas and dust toward core A appears to be at a higher
temperature than core B and most of the molecular emission lines of 
high energy transitions have been detected exclusively toward core A.  The measured
dust temperature toward core B is $\sim$40 K (Mundy et al.\ 1986) and the measured
temperature toward core A from many different molecular species is between 80 and 200 K
(Chandler et al.\ 2005).

Spectral observations with the IRAM 30-m radio telescope toward the low-mass 
protostellar system IRAS 16293-2422 (Cazaux et al.\ 2003) demonstrated an extremely rich organic 
inventory with abundant amounts of complex oxygen and nitrogen bearing molecules including formic acid 
(HCOOH), methyl formate (CH$_3$OCHO), acetic acid (CH$_3$COOH) and methyl cyanide (CH$_3$CN)
which are archetypal species often found in massive hot molecular cores (HMCs).  
Subsequent spatial observations by Kuan et al.\ (2004) with the Submillimeter Array (SMA) 
mapped emission from high energy transitions of CH$_3$OCHO, vibrationally excited
vinyl cyanide (CH$_2$CHCN) and other large molecular species. Additionally, Bottinelli et al.\ 
(2004) mapped slightly lower energy transitions of CH$_3$CN and CH$_3$OCHO with the Plateau 
de Bure interferometer, noting that all the molecular emission is contained toward cores A 
and B and that no large molecule emission is seen associated with the molecular outflows.  Recently, 
Chandler et al.\ (2005) detected large abundances of a number of different organic molecular
species primarily located toward core A from SMA observations.  Moreover, from observations of 
sulfur monoxide (SO) absorbing against the background dust continuum near binary core B, Chandler
et al.\ (2005) suggest that there must be material from an extended molecular envelope 
falling onto an embedded dust disk and that there may be a low-velocity 
outflow coming from an embedded protostar.

The discovery of large oxygen and nitrogen bearing molecules in a low-mass star forming region is significant
because the molecular complexity appears to be similar to high-mass star forming regions. Such
complex molecules undoubtedly are incorporated into subsequent formation of 
protoplanetary bodies and would provide a rich organic inventory that could be supplied 
to early planetary systems similar to that expected to have occurred in our own pre-solar nebula.  
With this in mind, we conducted high resolution observations of IRAS 16293-2422 using the 
NRAO\footnote{The National Radio Astronomy Observatory is a facility of
the National Science Foundation, operated under cooperative agreement by  Associated
Universities, Inc.} Very Large Array (VLA) using four molecular probes: ethyl cyanide (CH$_3$CH$_2$CN),
methyl formate (CH$_3$OCHO), formic acid (HCOOH) and the ground vibrational state of silicon monoxide (SiO) to determine
the implication of their relative spatial locations.  Surprisingly, the location of SiO is
not co-spatial with either core A or B.  This prompted us to process archival BIMA array $^{12}$CO data
for an explanation.

\section{OBSERVATIONS}

Observations of IRAS 16293-2422 were conducted with the VLA in 2005 June 27 and 28 
in its BnC configuration.  Formic acid (HCOOH) and methyl formate (CH$_3$OCHO) were
observed simultaneously in one correlator setting and ethyl cyanide (CH$_3$CH$_2$CN)
and silicon monoxide (SiO) in another correlator setting.  Table 1 lists the molecule (col. [1]),
transition quantum numbers (col. [2]), rest frequency (col. [3]), upper energy level
E$_u$ (col. [4]), line strength S (col. [5]), dipole moment $\mu_a$  (col. [6]), and molecular
parameter reference (col. [7]).  The phase center for all
observations was $\alpha$= 16$^h$32$^m$22$^s$.8, $\delta$ = -24$^o$28$'$33$''$ (J2000.0).
The correlator was operated in the four intermediate-frequency (IF) normal mode, and the 
selection of a 6.25 MHz bandpass yielded 32 spectral channels in two pairs of oppositely
polarized IFs giving a channel width of 195.313 kHz ($\sim$1.35 km s$^{-1}$ at 44 GHz).

The compact radio source 1625-254 was used as the primary phase calibrator.  The absolute
flux density scale was determined from observations of 1331+305 (3c286) whose average
flux density from the 2 observing bands was 1.44 Jy as given by the SETJY routine
in AIPS.  The averaged bootstrapped flux density of 1625-254 was 0.6 Jy.  Antenna
gains were derived from 1625-254 observations at 5 minute intervals.  After $\sim$1 hr,
a pointing calibration was performed to ensure pointing accuracy.  All data were combined
and calibrated using the AIPS data reduction package of NRAO.  The calibrated $u-v$ datasets
were then imported into MIRIAD (Sault et al.\ 1995) for imaging and deconvolution.  Figures
1-4 show the images of the emission from CH$_3$CH$_2$CN, CH$_3$OCHO, HCOOH,
and SiO, respectively toward IRAS 16293-2422.

The naturally weighted synthesized beamwidth resulting from the BnC configuration at 44 GHz
was 0.$''$4$\times$0.$''$3 and the largest angular size that can be measured in the
BnC array is 43$''$.  In order to determine the location and morphology
of the molecular emission toward 16293-2422, the calibrated $u-v$
data were INVERTed using synthesized beams ranging from the naturally weighted synthesized
beamwidth to a beamwidth of 20$''$$\times$20$''$.  This process indicated 
whether the bulk of the emission from a given species was compact or extended and was 
utilized since: (1) emission from HCOOH, SiO ($J=1-0$, $v$=0) and CH$_3$CH$_2$CN had not been previously
mapped toward this region, and (2) no emission feature of low energy transitions (E$_u$$<$7.5 K) 
of large molecules had been imaged toward this region.  

\section{RESULTS AND DISCUSSION}

\subsection{CH$_3$CH$_2$CN}

Figure 1 shows a contour map of the 5$_{15}$-4$_{14}$ transition of CH$_3$CH$_2$CN
superimposed on the 0.7 cm continuum emission (grayscale).  The mapped
CH$_3$CH$_2$CN emission region shows two blended emission peaks.  Compared
to the small naturally weighted beamwidth, a relatively large
synthesized beamwidth of 10.$''$0$\times$7.$''$5 is necessary to detect
CH$_3$CH$_2$CN which is extended with respect to cores A and B and the
oxygen bearing species (see $\S$ 3.2).  Figure 1 also shows a
hanning smoothed emission spectrum, which was extracted from the entire
region of mapped emission, with an rms noise level of $\sim$7 mJy beam$^{-1}$ shown on the left.  
The spectrum shows a relatively strong emission component at an LSR velocity of +7.7 km s$^{-1}$ and a possible
weaker component at +2.0 km s$^{-1}$.  The systemic LSR velocity of +3.9 km
s$^{-1}$ (Bottinelli et al.\ 2004) is shown as a dashed line.  We determine the total
beam-averaged CH$_3$CH$_2$CN column density, $N_{T}$, using the formula:

\begin{equation}
N_{T} = 2.04\frac{Q(T_{rot})e^{(E_u/T_{rot})} \int \Delta I dv }{\theta_a \theta_b \nu^3 S_{ij}\mu^2 }\times10^{20} \rm{cm^{-2}},
\end{equation}

\noindent
which is described in Remijan et al.\ (2004) where $E_u$ is the upper state energy level (K);
$T_{rot}$ is the rotational temperature (K); $\theta_a$ and $\theta_b$ are the major
and minor axes of the synthesized beam (arcsec); $\int \Delta I dv$ is the observed integrated line
intensity (Jy beam$^{-1}$ km s$^{-1}$); $\nu$ is the rest frequency (GHz); $S_{ij}\mu^2$ is the
product of the transition line strength and the square
of the electric dipole moment (Debye$^2$) and $Q(T_{rot})$ is the
rotational partition function. $N_{T}$(EtCN) $\sim$ 5.2$\times$10$^{14}$ cm$^{-2}$, 
which assumes a rotational temperature of 54 K (Cazaux et al.\ 2003), is listed in Table 2.  For a
molecular hydrogen column density of 7.5$\times$10$^{22}$ cm$^{-2}$ (Cazaux et al.\
2003), the CH$_3$CH$_2$CN fractional abundance, $X$(EtCN), is 6.9$\times$10$^{-9}$.
These values are consistent to the column density of CH$_3$CH$_2$CN previously measured
toward this source and is typical of the fractional abundance seen toward the Orion Compact 
Ridge (Cazaux et al.\ 2003).

The Figure 1 detection of extended emission with a spatial scale of
$\sim$20$"$ in a low-energy transition of CH$_3$CH$_2$CN demonstrates that there is
a large envelope of cold molecular gas surrounding the compact binary
cores A and B.  Core A is reputed to be hot (Chandler et al.\ 2005) while core B is
thought to be much cooler (Mundy et al.\ 1986).  A warm gas phase chemistry or
grain surface chemistry can lead to enhanced emission from nitrogen bearing molecules in hot molecular
cores (Remijan et al.\ 2004; Mehringer \& Snyder 1996).  In previous
observations of IRAS 16293-2422 which sample high energy level
transitions at 1 and 3 mm, enhanced emission of nitrogen bearing
molecules have been detected toward cores A and B (e.g., CH$_3$CN by 
Bottinelli et al.\ 2004; HCN, HC$_3$N, and CH$_2$CHCN by Kuan et al.\ 2004) 
on a typical spatial scale of $\sim$5$''$.  Thus, our
low-energy CH$_3$CH$_2$CN observations are complementary to these
high-energy observations of other nitrogen bearing molecules,
suggesting both increasing temperature and density gradients toward
cores A and B.

\subsection{CH$_3$OCHO and HCOOH}

High energy transition of a number of different molecules have
emission centroids near core A (e.g., see Figure 9 of Chandler et
al. 2005); in particular, CH$_3$OCHO has high energy transitions that are
associated with core A (Chandler et al.\ 2005; Bottinelli et al.\
2004; Kuan et al.\ 2004).  The only large oxygen bearing molecule detected
by Chandler et al.\ (2005) toward core B is a low energy transition
of dimethyl ether (CH$_3$OCH$_3$).  Figures 2 and 3 show maps of low-energy transitions
of CH$_3$OCHO and HCOOH, respectively, that are in proximity to core
B.  This is further evidence that core B is a much cooler
protostellar core than core A.  In Figure 2, the averaged emission
from both the A and E states of CH$_3$OCHO are mapped using a
synthesized beamwidth of $\sim$2.$''$6$\times$1.$''$8. 
Figure 2 also shows an emission complex centered on the A and E states if one
assumes an LSR velocity of $\sim$+3.9 km s$^{-1}$.  This complex
appears to be suffering from the competing effects of emission and
self-absorption.  In order to observe both states in the same
bandpass, the average frequency (45.3966 GHz) of
both states was used in the observations and this average
frequency corresponds to an LSR velocity of +3.9 km s$^{-1}$ in
Figure 2 while the A and E states appear at -1.6 and +9.4 km s$^{-1}$, 
respectively.  Figure 3 shows a map of the low-energy transition of HCOOH that surrounds the peak of
core B and also the CH$_3$OCHO emission in Figure 2; no HCOOH
emission was detected in the vicinity of core A.  In Figure 3, the emission
from HCOOH was mapped using a synthesized beamwidth of $\sim$2.$''$6$\times$1.$''$8; similar
to the beamwidth used to map the CH$_4$OCHO emission. The LSR
velocity of HCOOH is $\sim$+3.9 km s$^{-1}$ as indicated by the vertical
dashed line in the spectrum also shown in Figure 3.  

The spatial distributions of CH$_3$OCHO and HCOOH toward core B of
IRAS 16293-2422 compare favorably to those same spatial
distributions observed toward the Orion Molecular Cloud (OMC-1)
compact ridge (Hollis et al.\ 2003; Liu et al.\ 2002).  For
example, in the OMC-1 compact ridge, Hollis et al.\ (2003)
observed HCOOH emission surrounding CH$_3$OCHO emission, and
suggested that the HCOOH emission delineates the leading edge of a
shock front as the outflow from source "I" interacts with the
quiescent ambient gas; the CH$_3$OCHO emission, which is closer to
source "I", then represents the post shock gas.  If this scenario
is correct for IRAS 16293-2422, then the corresponding outflow
source associated with Figures 2 and 3 is likely core B itself. 
Indeed, Chandler et al.\ (2005) report evidence for material infall
toward and also low-velocity outflow from core B; both conditions
undoubtedly result in shock phenomena.  Toward core A, enhanced
abundances of large molecules provide indirect evidence for
shocks to explain the enhancements (Chandler et al.\ 2005).  The molecular
differences between cores A and B are most likely explained by
temperature differences and different protostellar ages.  In any
case, the presence of shocks seemingly leads to abundance
enhancements in oxygen-bearing molecules, including large
interstellar aldehydes located in high-mass star forming regions
like Sgr B2 (e.g., see Hollis et al.\ 2004a, b).

Because the CH$_3$OCHO emission appears to be self-absorbed, we can only give a lower limit to the 
column density over the entire emission complex. Since we have observed only low energy transitions in this
work, a conservative rotational temperature of 40 K (Mundy et al.\ 1986) is assumed 
which yields $N_T$(MeF) $>$1.9$\times$10$^{16}$ cm$^{-2}$. Assuming an H$_2$ 
column density of 1.6$\times$10$^{24}$ cm$^{-2}$ toward core B (Kuan et al.\ 2004), the 
CH$_3$OCHO fractional abundance is $X$(MeF) $>$1.2$\times$10$^{-8}$.  For the same 
rotational temperature, $N_T$(HCOOH) $\sim$ 4.1$\times$10$^{15}$ cm$^{-2}$ and the HCOOH 
fractional abundance is $X$(HCOOH) $\sim$ 2.5$\times$10$^{-9}$ (see Table 2). 

\subsection{SiO}

Figure 4 shows redshifted and blueshifted SiO ($J=1-0$, $v$=0) emission contours
mapped separately over the 0.7 cm continuum (grayscale).  These two components of the SiO emission are 
seen at LSR velocities of +2.3 km s$^{-1}$ and +7.9 km s$^{-1}$ 
which are approximately symmetrical around the systemic LSR velocity of 3.9 km s$^{-1}$. The peak 
intensity maps, which are reminiscent of SiO maser spots, are shown at these velocities.   The 
resulting synthesized beamwidth of $\sim$1.$''$5$\times$0.$''$8 is shown at the bottom 
left of the map. Also shown in Figure 4 is a spectrum which is averaged over the SiO emitting 
regions and hanning smoothed over 3 channels; the rms noise level is $\sim$3 mJy beam$^{-1}$ shown to the left.

This is the first detection of the $J=1-0$, $v$=0 transition of SiO toward IRAS 16293-2422 and the 
first map of the compact emission from SiO in the vicinity of cores A and B. Previous surveys by
Blake et al.\ (1994) and Ceccarelli et al.\ (2000) detected weak emission features of higher
energy transitions of $v$=0 SiO using single element radio telescopes.  From a chemical model of the IRAS 16293-2422
region, Ceccarelli et al.\ (2000) concluded that an enhanced abundance of SiO was necessary
to account for the excess emission from the higher $J$ level transitions and
that the emission must be coming from a compact ($<$3$''$) region.  Finally, their model
calculated an infall velocity of SiO of 2.8 km s$^{-1}$ which was consistent with their
observed $\sim$5 km s$^{-1}$ linewidths.  

From our high resolution VLA observations, we find that the blue and redshifted SiO 
emission peaks have a spatial separation 
of $\sim$2$''$ and the center of the SiO emission centroid is located at $\alpha$= 16$^h$32$^m$22$^s$.875, 
$\delta$ = -24$^o$28$'$32$''$.48 (J2000.0).  This location is within 0.$''$6
 of the pointing position and 
location of the $^{13}$CO emission centroid taken with the Owens Valley Radio Observatory (OVRO) millimeter 
array (Mundy et al.\ 1986) and the peak position of CS emission taken with the IRAM 30-m (Menten et al.\ 1987). 
From the red and blueshifted emission peaks, Mundy et al.\ (1986) inferred a rotational velocity of 2 km s$^{-1}$
and a 2 M$_{\odot}$ rotating disk of material internal to the $^{13}$CO emission.  From the CS data,
Menten et al.\ (1987) inferred a rotational velocity of 1.1 km s$^{-1}$ from lower resolution observations
a 4 M$_{\odot}$ rotating disk of material.  From the high resolution SiO data, we determine a 
rotational velocity of $\sim$2.8 km s$^{-1}$ and for a IRAS 16293-2422 distance of 160 pc, the mass interior
to the SiO emission is $\sim$1.4M$_{\odot}$.  However, assuming the SiO emission is associated with rotation, 
the compact disk containing the SiO emission is {\it counter-rotating} with respect to the
previously reported  $^{13}$CO and CS emission.  From archival BIMA array data taken of the $J=2-1$ 
transition of $^{12}$CO at 230.538 GHz, we find both emission and absorption toward IRAS 16293-2422.  
Moreover, the spectrum of the $^{12}$CO data clearly show an inverse P Cygni profile with the strongest 
absorption in close proximity to the SiO emission (Figure 5), indicating unambiguously, material infalling 
toward the molecular disk. Chandler et al.\ (2005) also suggest there is evidence for infall toward core B 
from the absorption profile of SO taken with the SMA.

It is important to note that we have only observed small-scale, blue-shifted emission and 
small-scale, red-shifted absorption of $^{12}$CO in the vicinity of cores A and B (see the image 
in Figure 5); these data possess the characteristic signature of infalling matter (i.e., an 
inverse P Cygni profile shown in Figure 5).  While the $^{12}$CO emission encompasses cores A and 
B, the more compact absorption occurs at a discrete position $\sim$3.$''$5 north of core A and 
$\sim$2.$''$5 west of core B (see Figure 4).  On the other hand, large-scale $^{12}$CO has been 
observed in red-shifted emission and blue-shifted emission by other investigators who 
reasonably interpret these results as outflow whose origin has traditionally been assigned to 
core A.  Thus, $^{12}$CO on two different spatial scales indicate two different phenomena:  one 
of infall on a small-scale toward a newly detected protostellar accretion disk whose 
position is precisely determined and one of large-scale outflow whose exact origin may be 
difficult to determine.  The location of the accretion disk that we have detected seems to 
account for our small-scale SiO observations and the somewhat larger-scale $^{13}$CO 
observations of Mundy et al.\ (1986) and CS observations of Menten et al.\ (1987) only if 
the nearly edge-on disk is counter-rotating since the direction of rotation is opposite for SiO 
when compared to either $^{13}$CO or CS.  Additional observational support for this scenario 
can be seen in the BIMA and OVRO array images of Sch\"{o}ier et al.\ (2005) who observed a 
number of low energy transitions (i.e., the J = 1 - 0 of HCO$^+$, HNC, and N$_2$H$^+$ and the $J = 
2 - 1$ of C$^{18}$O and SiO). In these images, for example, the $J = 2 - 1$ transition of SiO 
suggests rotation in the opposite direction to the  $J = 1 - 0$ transition of HCO$^+$.  Compared 
to the VLA data presented in this work, these BIMA and OVRO data have an order of 
magnitude coarser spatial resolution that precludes an exact determination of the SiO 
emission locations that are apparent in the VLA map shown in Figure 4.  Furthermore, SiO is a refractory 
molecule that can withstand high temperatures, and would likely exist closer to a forming 
protostellar source than most other molecules.  
Thus, the high resolution VLA observations have shown for the first time, evidence for a new
protostellar core with a compact inner portion of the disk 
containing the SiO emission counter-rotating with respect to the outer portion of the disk 
containing $^{13}$CO and CS.  Counter-rotating cores have been reported in disks of galaxies 
before (e.g., Thomas et al.\ 2005 and references therein), but this is the first indication of 
such behavior in a protostellar accretion disk. 

\section{CONCLUSIONS}

To summarize, we find a low temperature envelope of molecular CH$_3$CH$_2$CN 
emission that is encompassing the IRAS 16293-2422 cores A and B.  Furthermore, 
low frequency, low energy transitions of large oxygen bearing molecules (i.e. HCOOH 
and CH$_3$OCHO) are seen toward core B where an outflow may be responsible for 
a low velocity shock that is leading to enhanced abundances of large oxygen 
bearing species.  Also,  the SiO emission shows two velocity
components with spatial scales less than  2$''$ and a velocity separation of $\sim$5.6 km s$^{-1}$.
We interpret the spatial position offset in red and blueshifted SiO emission as due to the rotation of a 
protostellar accretion disk and we derive $\sim$1.4 M$_{\odot}$ interior to the SiO emission.

While it could be that the SiO complex we observe is due to outflow,
evidence is accumulating to the contrary that rotation is the
explanation.  Such evidence is best seen on a small scale.  The 
SiO observations in the present work show two velocity
components each with linewidths of $\sim$5 km s$^{-1}$, suggesting rotation of a
disk seen edge-on since the two components are not cospatial, are
symmetric about an infall location, and have a velocity gradient
perpendicular to large-scale CO outflow.  Moreover, images of other
molecules on a small-scale also show that their velocity gradients are
orthogonal to the direction of outflow seen on a large scale (e.g.,
Sch\"{o}ier et al.\ 2005).  In this work, we have been very careful to
match the synthesized beam to the source emission or absorption to
pin-point the location of a new source of infall $\sim$3.$''$5 north of core
source A and $\sim$2.$''$5 west of core source B.  We have done this by 
two independent sets of observations (i.e., the VLA in SiO emission and the BIMA array in $^{12}$CO absorption).
While there are no previous reports of continuum emission at this
location, an inverse P Cygni profile is absolute proof of absorption
against background continuum, and our BIMA data show that the location
is smaller than the synthesized beam (i.e., 3.2$''$$\times$0.8$''$).  Thus, the
continuum source is undoubtedly small (e.g. a disk seen edge-on) and
may eventually be detected by an interferometer at high resolution
with enough integration time.

However, the compact disk containing the SiO emission is {\it counter-rotating} with 
respect to the $^{13}$CO and CS emission seen by other investigators. This is the 
first report of evidence for a counter-rotating accretion disk toward a low-mass protostellar complex.
Moreover, archival BIMA array $^{12}$CO data show an inverse P Cygni profile with the 
strongest absorption in close proximity to the SiO emission, indicating unambiguously, material infalling
toward the counter-rotating protostellar disk.  The infall location is compelling evidence
for a new protostellar source within the IRAS 16293-2422 complex.

We thank F. Sch\"{o}ier and C. Ceccarelli for valuable comments on this work, and R.
Crutcher for permission to utilize the BIMA array data archive to
explore $^{12}$CO data toward IRAS 16293-2422. J.M.H.\ gratefully acknowledges 
research support from H.A.\ Thronson while assigned to the NASA Science Mission Directorate.

\begin{center}
{\bf FIGURE CAPTIONS}
\end{center}

\figcaption{Spectrum and map of the 5$_{15}$-4$_{14}$ transition of CH$_3$CH$_2$CN. 
The spectrum, which was taken from the entire region of mapped emission, is hanning smoothed over
3 channels.  The rms noise level of $\sim$7 mJy beam$^{-1}$ is shown at the left.  The systemic LSR velocity of +3.9 km
s$^{-1}$ is shown as a vertical dashed line.  The contour map of the 5$_{15}$-4$_{14}$ transition 
of CH$_3$CH$_2$CN is shown superimposed on 0.7 cm continuum emission (grayscale). The contour levels
are -0.045, 0.045, 0.060, 0.075 and 0.09 Jy beam$^{-1}$ (0.045 Jy beam$^{-1}$ = 5 $\sigma$). The synthesized beamwidth is shown 
at the bottom left of the map.}

\figcaption{Spectrum and map of the 5$_{15}$-4$_{14}$ A and E state transitions of CH$_3$OCHO. 
The spectrum, which was taken from the entire region of mapped emission, is hanning smoothed over
3 channels.  The rms noise level of $\sim$2 mJy beam$^{-1}$ is shown at the left.
The average frequency of the A and E state transitions appears at an LSR velocity of +3.9 km s$^{-1}$.
The contour map of the CH$_3$OCHO emission is shown superimposed on 0.7 cm continuum 
emission (grayscale). The contour levels are -0.008, -0.010, 0.010, 0.012, 0.014 and 0.016 Jy beam$^{-1}$
(0.010 Jy beam$^{-1}$ = 5 $\sigma$).  The synthesized beamwidth is shown at the bottom left of the map.}

\figcaption{Spectrum and map of the 2$_{02}$-1$_{01}$ transition of HCOOH. 
The spectrum, which was taken from the southernmost region of mapped emission 
toward core B, is hanning smoothed over 3 channels.  The rms noise level of 
$\sim$2 mJy beam$^{-1}$ is shown at the left. The spectrum shows the HCOOH 
emission is centered at the systemic LSR velocity of +3.9 km s$^{-1}$ (vertical dashed line). 
The contour map of the HCOOH emission is shown superimposed on 0.7 cm continuum 
emission (grayscale).  The contour levels are -0.007, -0.009, 0.011, 0.014, 0.016 and 0.018 Jy beam$^{-1}$ 
(0.011 Jy beam$^{-1}$ = 5 $\sigma$).  The synthesized beamwidth is shown at the bottom left of the map.}

\figcaption{Spectrum and maps of the $J$=1-0, $v$=0 transition of SiO. The spectrum of SiO was averaged 
over the emission regions and is hanning smoothed over 3 channels. The spectrum shows a redshifted and 
blueshifted components of SiO emission approximately centered at the systemic LSR velocity of 
+3.9 km s$^{-1}$ (vertical dashed line).  The 1 $\sigma$ rms noise level of $\sim$2 mJy beam$^{-1}$ is 
shown on the left. The contour map of the SiO emission is superimposed on the 0.7 cm continuum 
emission (grayscale) and shows the location of the redshifted and blueshifted emission peaks.  
The contour levels are -0.008, -0.010, 0.010, 0.011, 0.012, 0.013 and 0.014 Jy beam$^{-1}$ 
(0.010 Jy beam$^{-1}$ = 5 $\sigma$). The synthesized beamwidth is shown at the 
bottom left of the map.  Also shown in the map is the location of the absorption trough of the inverse 
P Cygni profile (filled circle) of the $^{12}$CO emission shown in Figure 5.}

\figcaption{Map and spectrum of the $J$=2-1 transition of $^{12}$CO at 230.538 GHz taken with the BIMA array. 
The contour map of the $^{12}$CO emission is superimposed on the 1 mm continuum 
emission (grayscale) and shows the location of the redshifted absorption and blueshifted emission.
The contour levels are -5.50, -5.00, -4.50, -4.00, -3.50, 2.25, 3.00, 3.75, 4.50, 5.25 and 6.00 Jy beam$^{-1}$. 
The synthesized beamwidth of 3.$''$2$\times$0.$''$8 is shown at the bottom left of the map. The spectrum 
of the $J$=2-1 transition of $^{12}$CO at 230.538 GHz was taken at a location surrounding the 
absorption and emission peaks and has a 1 $\sigma$ rms noise level of $\sim$1 Jy beam$^{-1}$ shown on the left of the 
spectrum.  The systemic LSR velocity of +3.9 km s$^{-1}$ is shown as a vertical dashed line.}

\begin{deluxetable}{ccccccc}
\tablewidth{34pc}
\tablecolumns{7}
\tablecaption{Molecular Line Parameters for VLA Observations}
\tablehead{
\colhead{} & \colhead{Transition} & \colhead{Frequency\tablenotemark{a}} & \colhead{E$_u$} &  \colhead{} & \colhead{$\mu_a$} & \colhead{} \\
\colhead{Molecule} & \colhead{$J'_{K-,K+}$-$J''_{K-,K+}$} & \colhead{(MHz)} & \colhead{(K)} & \colhead{S} & \colhead{(D)} & \colhead{Reference}\\
\colhead{(1)} & \colhead{(2)} & \colhead{(3)} & \colhead{(4)} & \colhead{(5)} & \colhead{(6)} & \colhead{(7)}\\
}
\startdata
SiO & 1-0,$v$=0 & 43,423.864(22) & 2.085 & 1.00 & 3.10 & b\\
CH$_3$CH$_2$CN & 5$_{15}$-4$_{14}$ & 43,516.200(8) & 7.380 & 4.80 & 3.85 & c\\
HCOOH & 2$_{02}$-1$_{01}$ & 44,911.750(50) & 3.238 & 2.00 & 1.39 & d\\
CH$_3$OCHO & 4$_{14}$-3$_{13}$ E & 45,395.755(16) & 6.187 & 3.80 & 1.63 & e\\
                            & 4$_{14}$-3$_{13}$ A & 45,397.394(16) & 6.187 & 3.80 & 1.63 & e\\
\enddata
\tablenotetext{a}{Uncertainties in parentheses refer to the least significant digit and are 2 $\sigma$ values.}
\tablenotetext{b}{(Lovas \& Krupenie 1974)}
\tablenotetext{c}{(Lovas 1982)}
\tablenotetext{d}{Willemot et al.\ 1980}
\tablenotetext{e}{Oesterling et al.\ 1999}
\end{deluxetable}

\begin{deluxetable}{cccccc}
\tablewidth{37pc}
\tablecolumns{7}
\tablecaption{Molecular Column Densities}
\tablehead{
\colhead{} & \colhead{$v_{LSR}$ Range} & \colhead{$\int$I dv} &  \colhead{T$_{rot}$} & \colhead{$N_{T}$} & \colhead{$X$} \\
\colhead{Molecule} & \colhead{(km s$^{-1}$)} & \colhead{(Jy bm$^{-1}$ km s$^{-1}$)} & \colhead{(K)} & \colhead{($\times$10$^{14}$ cm$^{-2}$)} & \colhead{($\times$10$^{-9}$)}\\
\colhead{(1)} & \colhead{(2)} & \colhead{(3)} & \colhead{(4)} & \colhead{(5)} & \colhead{(6)} \\
}
\startdata
CH$_3$CH$_2$CN & +0.0 to +10.0 & 0.34(9) & 54 & 5.2(1.4) & 6.9(1.4)\\
CH$_3$OCHO & -5.0 to +16.0 & $>$0.22 & 40 & $>$187 & $>$11.7\\
HCOOH & +0.0 to +8.0 & 0.07(3) & 40 & 41(17) & 2.5(1.7)\\
\enddata

\tablecomments{SiO is likely masering and is therefore not included in this table (see text).}
\end{deluxetable}

\begin{figure}
\epsscale{0.65}
\plotone{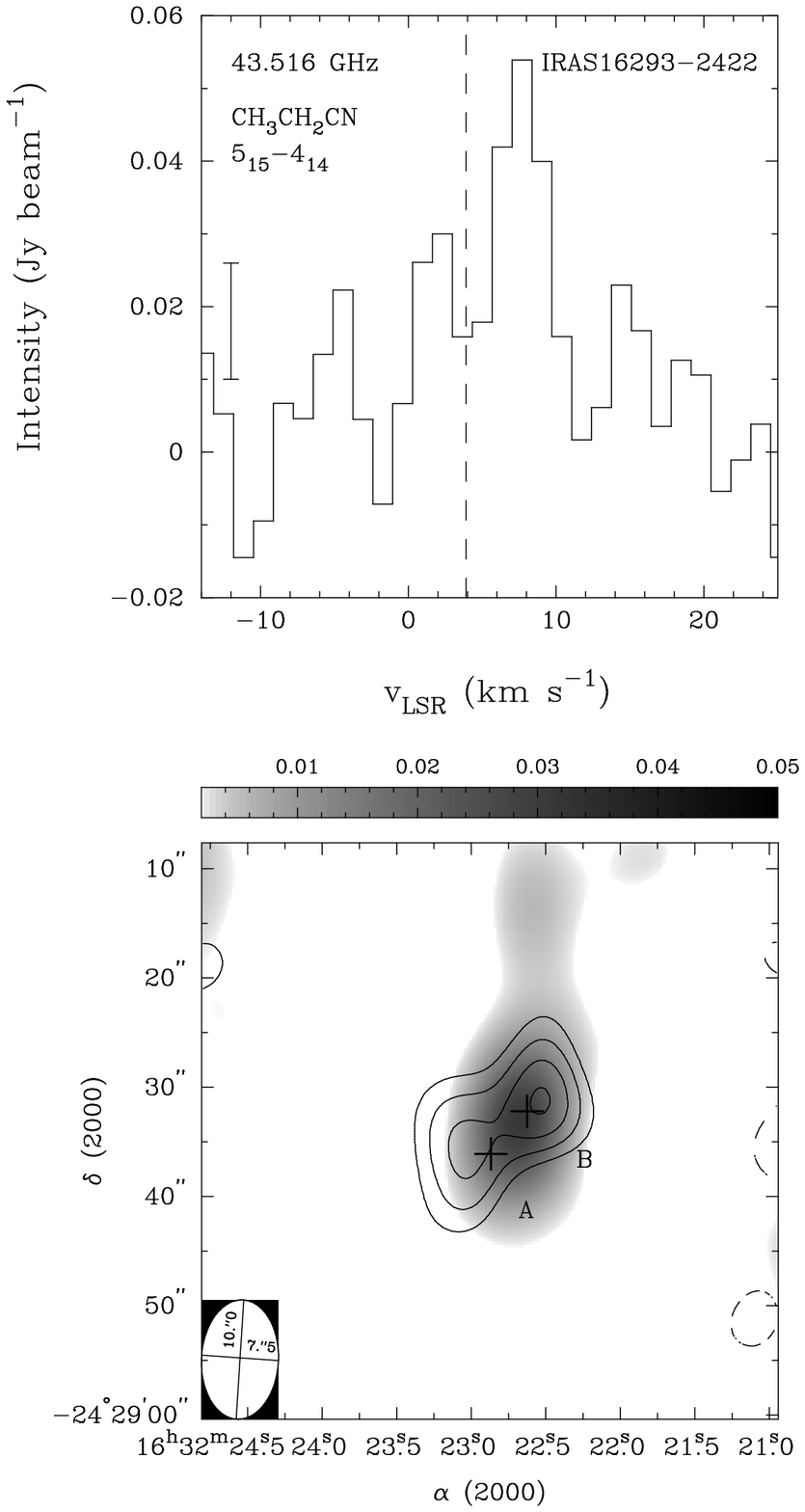}
\centerline{Figure 1.}
\end{figure}

\begin{figure}
\epsscale{0.65}
\plotone{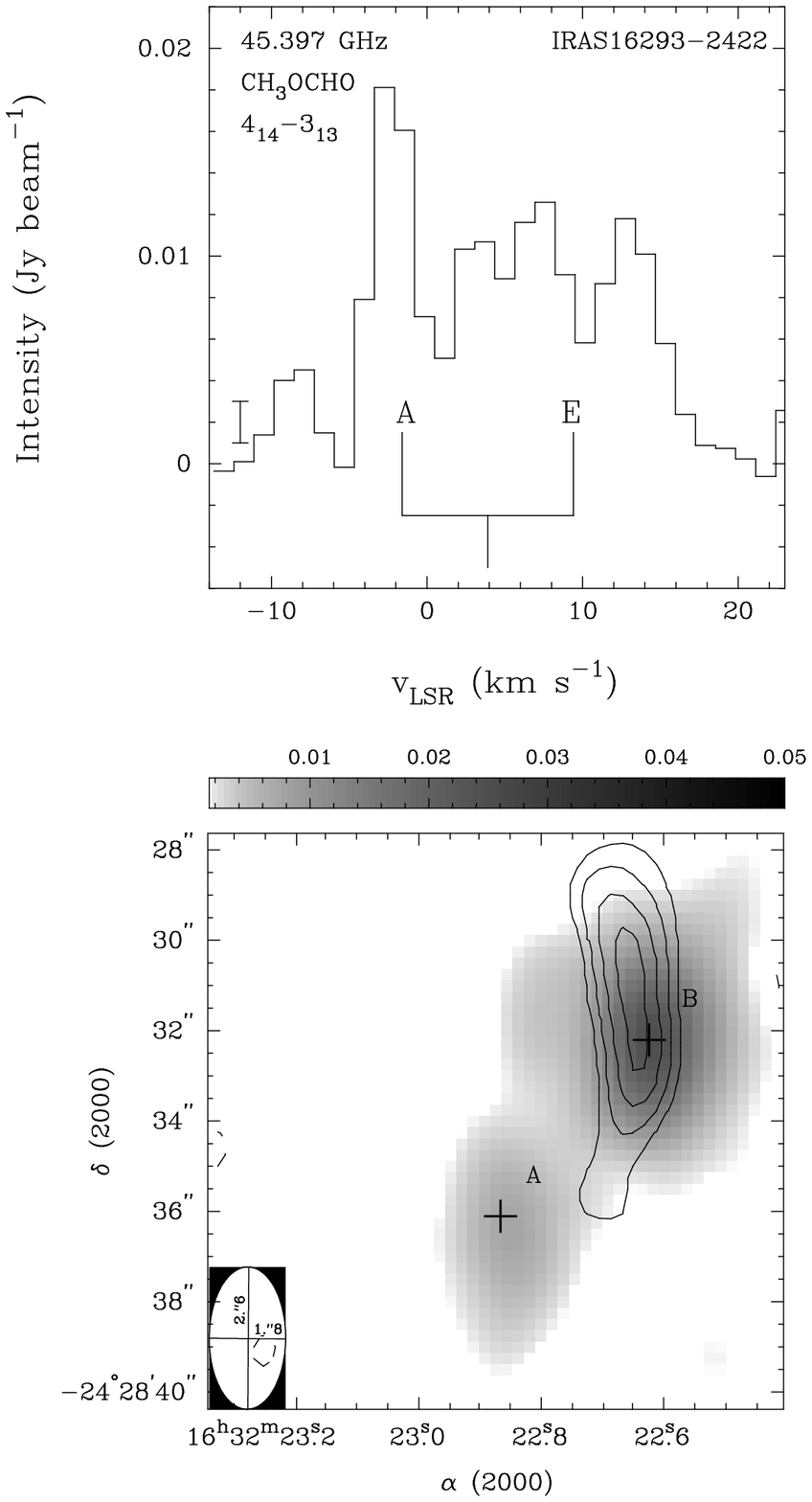}
\centerline{Figure 2.}
\end{figure}

\begin{figure}
\epsscale{0.650}
\plotone{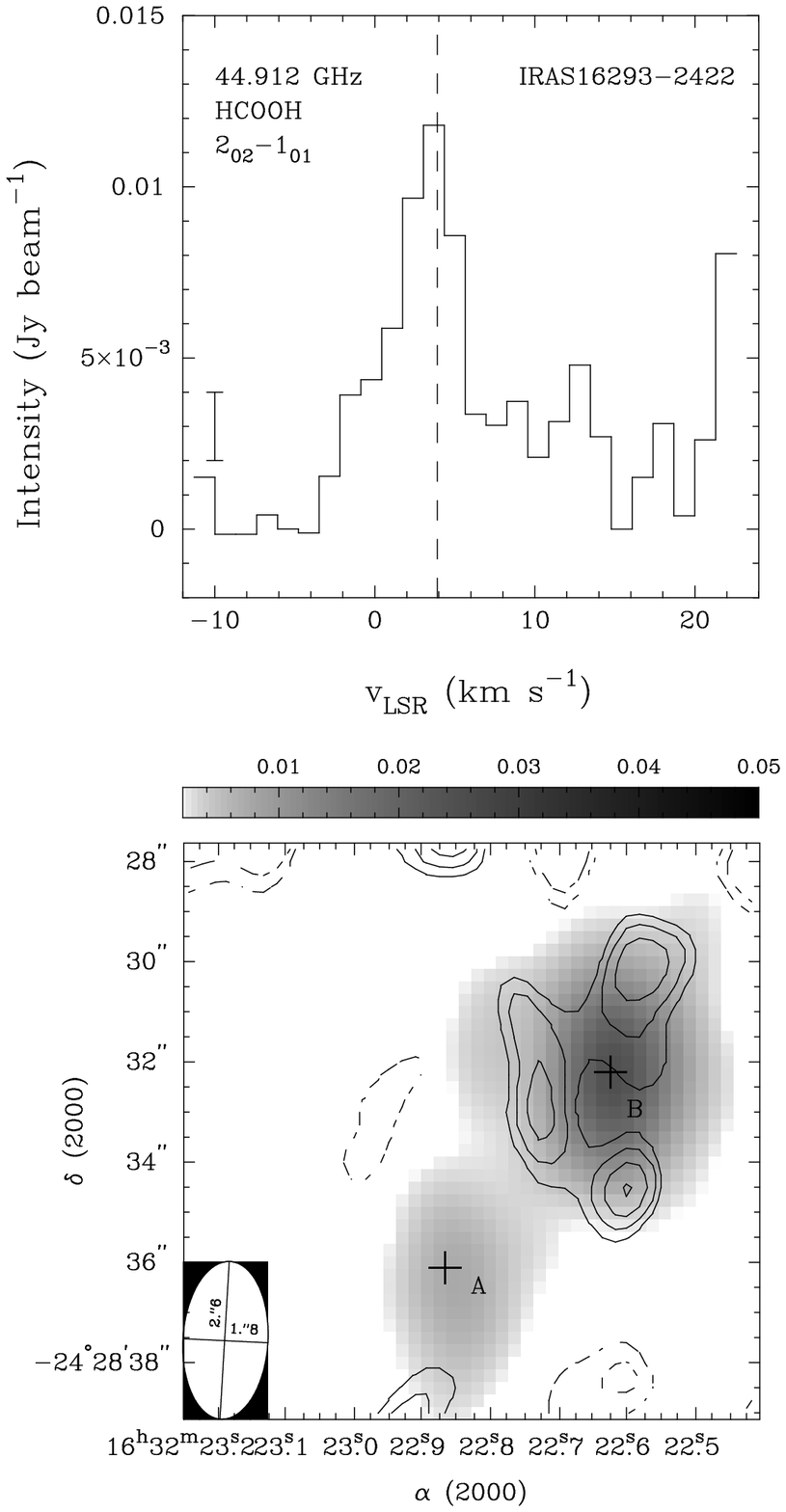}
\centerline{Figure 3.}
\end{figure}

\begin{figure}
\epsscale{0.9}
\plotone{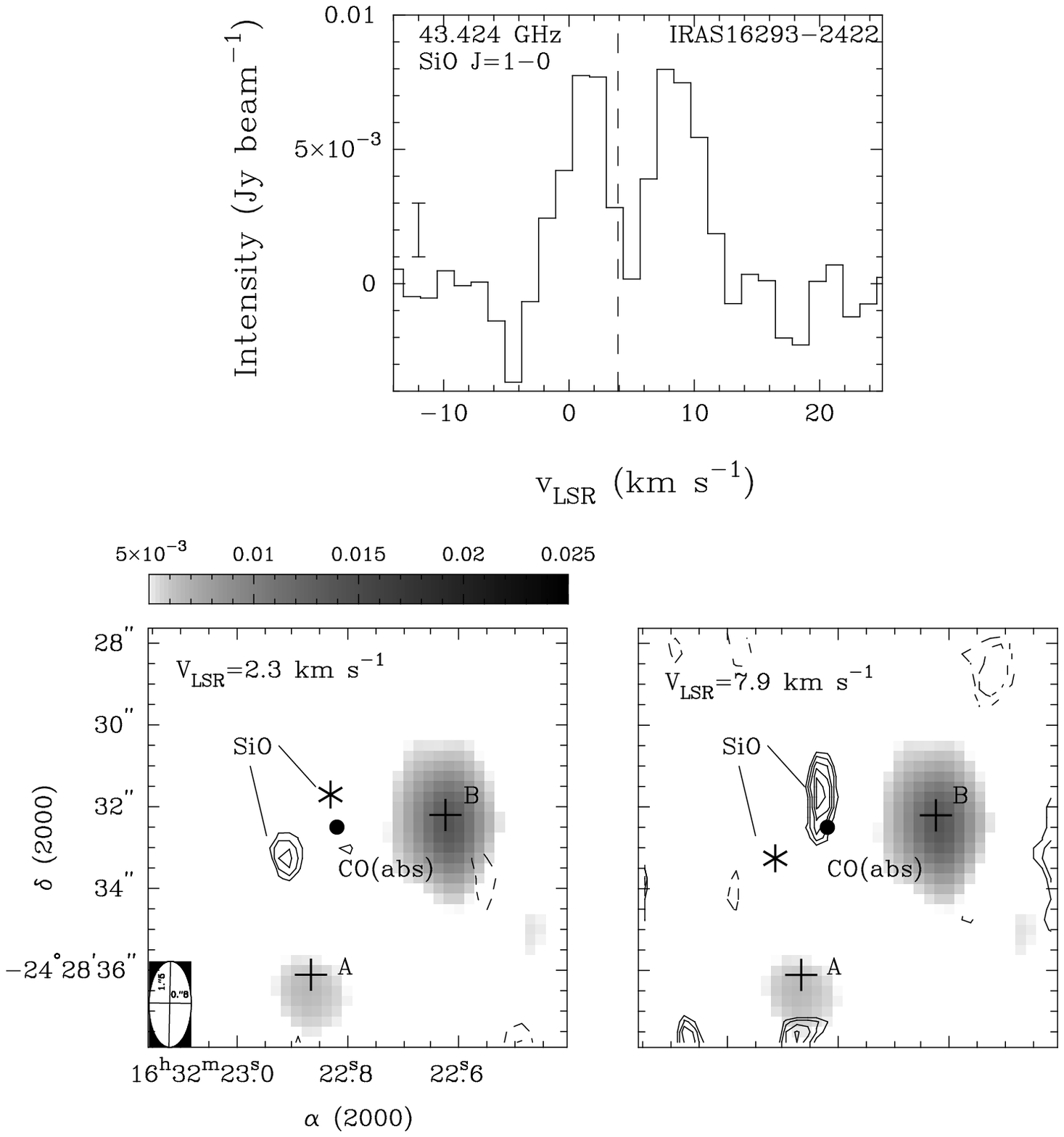}
\centerline{Figure 4.}
\end{figure}

\begin{figure}
\epsscale{0.65}
\plotone{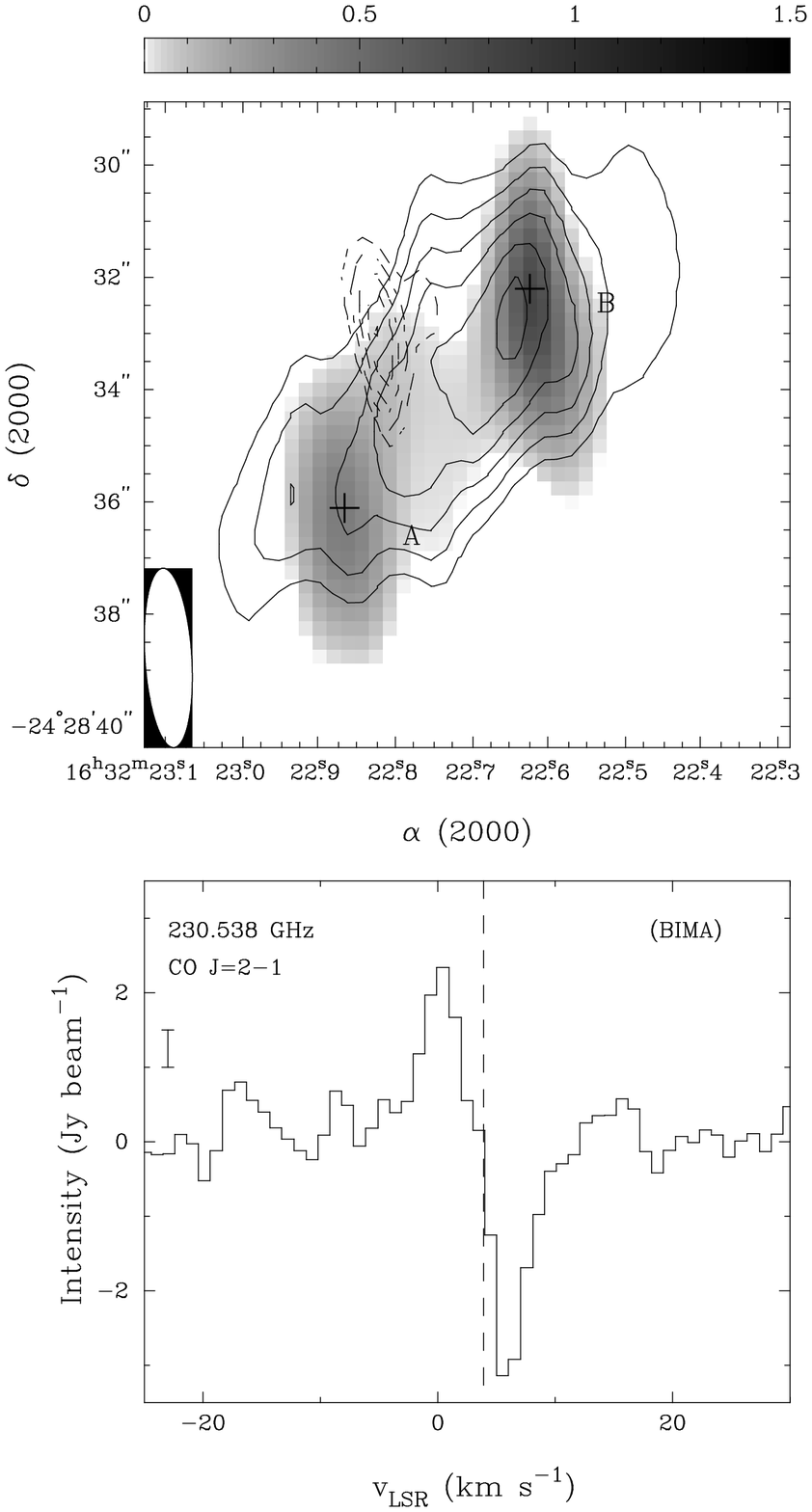}
\centerline{Figure 5.}
\end{figure}

\end{document}